\documentclass[12pt,epsf]{article}
\usepackage{graphicx}
\usepackage{amssymb}

\begin{document}

\def\a{\alpha}
\def\b{\beta}
\def\c{\varepsilon}
\def\d{\delta}
\def\e{\epsilon}
\def\f{\phi}
\def\g{\gamma}
\def\h{\theta}
\def\k{\kappa}
\def\l{\lambda}
\def\m{\mu}
\def\n{\nu}
\def\p{\psi}
\def\q{\partial}
\def\r{\rho}
\def\s{\sigma}
\def\t{\tau}
\def\u{\upsilon}
\def\v{\varphi}
\def\w{\omega}
\def\x{\xi}
\def\y{\eta}
\def\z{\zeta}
\def\D{\Delta}
\def\G{\Gamma}
\def\H{\Theta}
\def\L{\Lambda}
\def\F{\Phi}
\def\P{\Psi}
\def\S{\Sigma}
\def\BR{{\rm Br}}
\def\o{\over}
\def\beq{\begin{eqnarray}}
\def\eeq{\end{eqnarray}}
\newcommand{\nn}{\nonumber \\}
\newcommand{\gsim}{ \mathop{}_{\textstyle \sim}^{\textstyle >} }
\newcommand{\lsim}{ \mathop{}_{\textstyle \sim}^{\textstyle <} }
\newcommand{\vev}[1]{ \left\langle {#1} \right\rangle }
\newcommand{\bra}[1]{ \langle {#1} | }
\newcommand{\ket}[1]{ | {#1} \rangle }
\newcommand{\EV}{ {\rm eV} }
\newcommand{\KEV}{ {\rm keV} }
\newcommand{\MEV}{ {\rm MeV} }
\newcommand{\GEV}{ {\rm GeV} }
\newcommand{\TEV}{ {\rm TeV} }
\def\diag{\mathop{\rm diag}\nolimits}
\def\Spin{\mathop{\rm Spin}}
\def\SO{\mathop{\rm SO}}
\def\O{\mathop{\rm O}}
\def\SU{\mathop{\rm SU}}
\def\U{\mathop{\rm U}}
\def\Sp{\mathop{\rm Sp}}
\def\SL{\mathop{\rm SL}}
\def\tr{\mathop{\rm tr}}

\newcommand{\bear}{\begin{array}}  
\newcommand {\eear}{\end{array}}
\newcommand{\la}{\left\langle}  
\newcommand{\ra}{\right\rangle}
\newcommand{\non}{\nonumber}  
\newcommand{\ds}{\displaystyle}
\newcommand{\red}{\textcolor{red}}
\def\ubl{U(1)$_{\rm B-L}$}
\def\REF#1{(\ref{#1})}
\def\lrf#1#2{ \left(\frac{#1}{#2}\right)}
\def\lrfp#1#2#3{ \left(\frac{#1}{#2} \right)^{#3}}
\def\OG#1{ {\cal O}(#1){\rm\,GeV}}

\def\TODO#1{ {\bf ($\clubsuit$ #1 $\clubsuit$)} }


\baselineskip 0.7cm

\begin{titlepage}

\begin{flushright}
CALT  68-2857 \\
UT-11-49
\end{flushright}

\vskip 1.35cm
\begin{center} 
{\large \bf Natural Supersymmetry and $b \to s \gamma$ constraints }
\vskip 1.2cm

{Koji Ishiwata$^{(a)}$, Natsumi Nagata$^{(b,c)}$ and Norimi Yokozaki$^{(c)}$}

\vskip 0.4cm

{\it
$^a$California Institute of Technology, Pasadena, CA 91125, USA\\
$^b$Department of Physics, Nagoya University, Nagoya 464-8602, Japan\\
$^c$Department of Physics, University of Tokyo, Tokyo 113-0033, Japan
}

\vskip 1.5cm

\abstract{ We investigate constraints from the observed branching
  ratio for $b \to s \gamma$ and fine-tuning in the framework of
  natural supersymmetry.  The natural supersymmetry requires the large
  trilinear coupling of the stop sector, light higgsinos (a small
  $\mu$ parameter) and light stops, in order to reduce the fine-tuning
  in the Higgs sector while avoiding the LEP constraint.  It is found
  that in such a scenario 5\% (10\%) level of fine-tuning is
  inevitable due to the $b \to s \gamma$ constraint even if the
  messenger scale is as low as $10^5$ GeV ($10^4$ GeV),  provided that the gaugino masses satisfy the GUT relation.  }
\end{center}
\end{titlepage}

\setcounter{page}{2}

\section{Introduction}
The Standard Model (SM) of particle physics has achieved a remarkable
success in describing physical phenomena accessible so far. In the SM,
Higgs boson plays an important role since it breaks the electroweak
symmetry and becomes the origin of the masses of the SM particles. The
Higgs mass itself, however, is not protected by any symmetry, and thus
receives disastrously large radiative corrections. It is so-called the
hierarchy problem, or the naturalness problem, which may indicate that
the SM is just a low-energy effective theory. One of the most
attractive extensions of the SM is the Minimal Supersymmetric Standard
Model (MSSM). Supersymmetry (SUSY) protects SM-like Higgs mass from
quadratic-divergent correction, and then it provides a good solution
to the hierarchy problem when superparticles exist just above the
electroweak scale.

The ATLAS and CMS collaborations are now searching the superparticles
at the Large Hadron Collider (LHC). Since there has been no signal of
superparticles so far, the collaborations provide stringent limits on
the masses of colored superparticles ~\cite{Aad:2011ib,
  Chatrchyan:2011zy}. However, these limits are directly applicable
only to gluino and the first generation squarks in the general
context. Naturalness in the supersymmetric models, on the other hand,
requires the third generation squarks, especially stops, to have their
masses around the electroweak scale. Thus, it is worth studying
the case where only the third generation squarks are relatively light
and the other superparticles are beyond the reach of the current
collider experiments.

When considering the light stop scenario, one has to evaluate the
Higgs mass carefully. This is because the prediction of the Higgs mass
in the scenario might be lower than the LEP-II bound, 114.4
GeV~\cite{Barate:2003sz}.  In the evaluation of the Higgs mass in the
MSSM, radiative corrections from top/stop loops are important. Since
the tree-level SM-like Higgs boson mass is lighter than the $Z$ boson
mass, the stop masses cannot be too light.  As listed in
Refs.~\cite{Kitano:2006gv, Asano:2010ut}, naturalness argument calls
for several conditions in addition to light stops; an adequate value
of $\tan\beta$ in the Higgs sector, the large stop trilinear coupling
$A_t$, the small Higgsino mass parameter $\mu$ in the superpotential
and the small messenger scale. (Definition of those parameters are
given in the next sction.) The first two conditions are related to the
Higgs mass bound~\cite{Kitano:2005wc,Dermisek:2006ey,
  Dermisek:2006qj}, while the others are to improve naturalness, which
is discussed in the subsequent section.  Although such a light stop
scenario is favored in terms of the naturalness argument,
contributions of superparticles to the inclusive decay rate $B\to X_s
\gamma$ are enhanced~\cite{Bertolini:1990if}, and it should be
carefully taken care of.

In this letter we study the level of fine-tuning by taking into
account the constraints from the branching ratio for $b\rightarrow s
\gamma$, as well as the LEP-II bound for the mass of the SM-like higgs
boson.  {Assuming the GUT relation among gaugino masses,} it is
found that such region is severely constrained, and consequently at
least about 5 \% fine-tuning is required within the framework of the
MSSM.\footnote{
In the following analysis, we set the messenger scale to be $10^5$
GeV.
When the messenger scale is $10^4$ GeV, the minimum
  requirement of fine-tuning is relaxed, up to 10\%. } 
Constraints on
natural supersymmetry is also discussed based on the recent LHC
results in Ref.~\cite{Papucci:2011wy}. (See also the earlier
  work~\cite{Perelstein:2007nx}.) While the constraints given there
would be relaxed by $R$-parity violation our result is applicable even for such a case.

\section{Naturalness}

In the MSSM, there are two Higgs fields, $H_u$ and $H_d$, with their
vacuum expectation values (VEVs), $v_u$ and $v_d$, breaking the
electroweak symmetry. The VEVs are corresponding to the minimum of the
Higgs potential, and in the radial direction to the minimum, the
potential is simply written as
\begin{eqnarray}
V = m^2 |h|^2 + \frac{\lambda}{4} |h|^4,
\label{potential}
\end{eqnarray}
where $h$ is a linear combination of the Higgs fields. The mass of the
physical Higgs boson $m_h$ is determined by the curvature of the
potential in the direction around the minimum. A brief calculation
leads to the relation
\begin{equation}
 m_h^2=-2m^2~.
\end{equation}
At tree-level, the physical Higgs mass is bounded above in the MSSM:
\begin{equation}
 m_h\le m_Z |\cos(2\beta)|,
\end{equation}
where $m_Z$ is the mass of $Z$ boson, and $\tan\beta \equiv v_u/v_d$.
The relation originates from the fact that the quartic coupling
$\lambda$ in Eq.~(\ref{potential}) is written in terms of the
electroweak gauge couplings.
The upper limit is satisfied in the so-called decoupling limit, where
the mass of the CP-odd neutral scalar $m_A$ is much larger than $m_Z$.
In this case, the lighter Higgs mass eigenstate behaves like the SM
Higgs boson.  However, even in the decoupling limit with a large value
of $\tan \beta$, the tree level mass is not sufficient to exceed the
LEP bound.
It is accomplished with the help of radiative
corrections~\cite{Okada:1990vk}.  In the region with (moderately)
large $\tan\beta$, $m^2$ in Eq.~(\ref{potential}) is expressed as
\begin{eqnarray}
m^2 \simeq |\mu|^2 +  \left. m_{H_u}^2\right|_{M_{\rm mess}} 
+ \delta m_{H_u}^2, 
\label{eq:mhiggs}
\end{eqnarray}
where $\mu$ is the mass for the Higgs superfields and
$\left. m_{H_u}\right|_{M_{\rm mess}}$ is the soft SUSY breaking mass
parameter for the up-type Higgs at a scale $M_{\rm mess}$ where the
soft SUSY breaking terms are generated. The third term on the
right-hand side is radiative correction from the renormalization group
running between $M_{\rm mess}$ and the stop mass scale
$m_{\tilde{t}}$, which can be estimated as~\footnote{Although this
  estimation is a leading log approximation, in numerical
  calculations, we evaluate $\delta m_{H_u}^2$ by solving
  renormalization group equations, with gluino contributions included.
}
\begin{eqnarray}
\delta m_{H_u}^2 
&\simeq&
\left. m^2_{H_u}\right|_{m_{\tilde{t}}}-\left. m^2_{H_u}\right|_{M_{\rm mess}}
\nonumber \\
&\simeq& 
-\frac{3 y_t^2}{8\pi^2} (m_{\tilde{Q}_3}^2 + m_{\tilde{t}_R}^2 + |A_t|^2)
\ln \frac{M_{\rm mess}}{m_{\tilde{t}}},
\end{eqnarray}
where $y_t$ is the top Yukawa coupling, and $A_t$ is the trilinear
scalar coupling of the top sector.  $m_{\tilde{Q}_3}^2$ and
$m_{\tilde{t}_R}^2$ denote the soft SUSY breaking mass parameters of
third generation squark doublet and singlet, respectively.  Since the
radiative correction is controlled by the mass scale of the stops, the
realization of the electroweak symmetry breaking requires fine-tuning
among three terms in Eq.~(\ref{eq:mhiggs}), unless $M_{\rm mess}$ is
sufficiently low and/or the soft masses for the stops are small.
In order to quantify the level of fine-tuning, we define a measure of
the fine-tuning as~\cite{Kitano:2006gv}
\begin{eqnarray}
\Delta^{-1} = \frac{m_h^2}{2 (- \delta m_{H_u}^2)} .
\end{eqnarray}
It is obvious that larger values of the stop mass parameters lead
to further fine-tuning.  Note that a large $\mu$ parameter also
requires the cancellation, therefore the maximum value is constrained
by
\begin{eqnarray}
|\mu| \lesssim 210~{\rm GeV} 
\left(\frac{15\%}{{\Delta}^{-1}}\right)^{1/2} 
\left(\frac{m_h}{115\, {\rm GeV}}\right). \label{eq:mu}
\end{eqnarray}
The Higgs mass bound given by the LEP II experiment, $m_h > 114.4 {\rm
  GeV}$, can be satisfied if one sets $A_t$ instead of the stop soft
masses to be large.  Therefore it is expected that the fine-tuning is
reduced when the stops are light and $A_t$ is large. However, we will
show that with such light stops and large $A_t$, the chargino
contribution to the $b \to s \gamma$ process becomes large and then
wide range of the parameter space where $\Delta^{-1} \gtrsim$ a few \%
is excluded.

\section{The constraint from $b \to s \gamma$}
The process $b \to s \gamma$ is suppressed by a loop-factor and the
off-diagonal element of the CKM matrix in the SM. The
experimental value of $b \to s \gamma$ roughly agrees with the
prediction in the SM:
\begin{eqnarray}
{\rm Br}(B \to X_s \gamma)_{\rm exp}&=&(3.55 \pm 0.26) \times 10^{-4},
\\
{\rm Br}(B \to X_s \gamma)_{\rm SM}&=&(3.15 \pm 0.23) \times 10^{-4}.
\end{eqnarray}
Here ``exp'' and ``SM'' mean the experimental
value~\cite{Asner:2010qj} and the SM prediction including
next-to-next-to-leading order QCD corrections~\cite{Misiak:2006zs} of
the branching ratio for the inclusive radiative decay $B \to X_s
\gamma$.  The deviation of the SM prediction from the observation is
\begin{eqnarray}
  -0.3 \times 10^{-4} < \Delta {\rm Br}(B \to X_s \gamma)
  < 1.1 \times 10^{-4},
\end{eqnarray}
at 2$\sigma$ level.  Generally, in supersymmetric models, the loop diagrams
in which superparticles run could induce comparable or even larger
contributions to $b \to s \gamma$~\cite{Bertolini:1990if}, resulting in
significant deviation from the SM prediction.  For example, it is
known that the light sparticles with large $\tan\beta$ are severely
constrained from the experimental data.

Important contributions to $b \to s \gamma$ arise from the charged
Higgs and chargino loop diagrams within the framework of the minimal
flavor violation in which the only source for flavor/CP violation
arises through the CKM matrix elements. (For SUSY contributions to
CP/flavor violating processes, see Ref.~\cite{Altmannshofer:2009ne}.)
The charged Higgs contribution always increases the branching ratio
since it interferes constructively with the SM contribution. On the
other hand, the chargino contributions can be either constructive or
destructive, depending on the sign and the size of the $\mu$
parameter, the wino mass, $M_2$, and the trilinear coupling
$A_t$. When both $M_2$ and $\mu$ are positive, a negative value of
$A_t$ decreases the branching ratio while a (sufficiently large)
positive $A_t$ increases it.\footnote{We follow the convention in
  the SLHA~\cite{Skands:2003cj}.}  In a nutshell, contributions from
superparticles in the loops to $b \to s \gamma$ are enhanced by a
small $\mu$ parameter, a large $A_t$, light stops and a large
$\tan\beta$.  However, avoiding the enhancement of $b \rightarrow s
\gamma$, a certain level of fine-tuning is expected in such a
parameter region, as discussed in the previous section.

In our numerical calculation, we demand that contributions of
superparticle loops to $b \to s \gamma$ should not exceed the
experimental value.
Namely, we constrain parameter region in the MSSM by imposing
\begin{eqnarray}
  -0.3 \times 10^{-4} < \Delta' {\rm Br}(B \to X_s \gamma)
  < 1.1 \times 10^{-4},
\label{eq:bsg}
\end{eqnarray} 
where $\Delta' {\rm Br}(B \to X_s \gamma)={\rm Br}'(B \to X_s
\gamma)_{\rm MSSM}-{\rm Br}'(B \to X_s \gamma)_{\rm SM}$. (Primes
indicate results from our numerical calculation ).

\begin{figure}[t]
\begin{center}
\includegraphics[width=9cm]{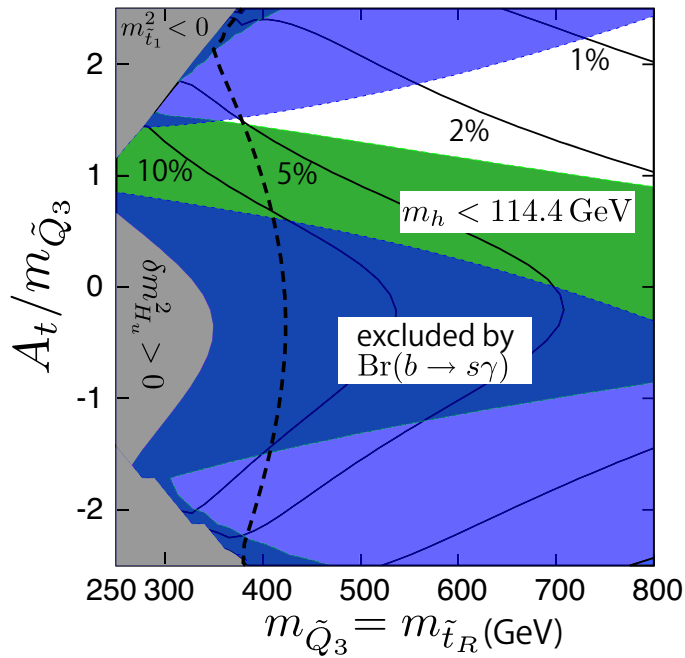}
\caption{Contours of the fine-tuning measure $\Delta^{-1}$. The SUSY
  parameters are set to $m_A=$1 TeV, $\tan\beta=10$ and $\mu=200$ GeV.
  The gluino mass is taken as $M_3=750$ GeV, and the GUT relation
  among gaugino masses is assumed so that $M_1:M_2:M_3 = 1:2:6$. The
  input parameters are defined at the scale of
  $m_{\tilde{Q}_3}=m_{\tilde{t}_R}$. 
 The messenger scale is set to be $10^5$ GeV.
 The blue region is excluded by $b
  \to s \gamma$ and the green region is excluded by the LEP bound.  In the left region of the black dashed line, $m_{\tilde{Q}_3}^2$ is negative at the messenger scale, $m_{\tilde{Q}_3}^2(M_{\rm mess})<0$.}
\label{fig:pos_mu}
\end{center}
\end{figure}

\begin{figure}[t]
\begin{center}
\includegraphics[width=15cm]{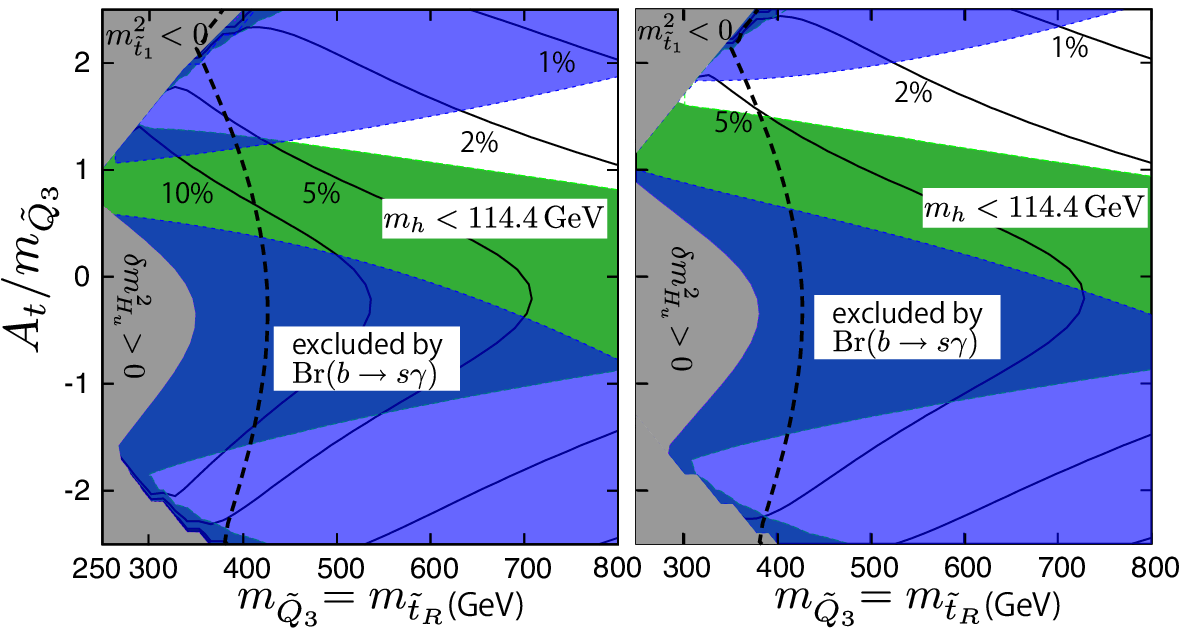}
\caption{Contours of the fine-tuning measure $\Delta^{-1}$. The $\mu$
  parameters are taken as $\mu=-200$ GeV in the left panel, and
  $\mu=300$ GeV in the right panel.  Other parameters are same as in
  figure \ref{fig:pos_mu}.  }
\label{fig:nega_mu}
\end{center}
\end{figure}
In Fig.~\ref{fig:pos_mu}, the contours of $\Delta^{-1}$ and the
constraint from the $b \to s \gamma$ are shown. The blue and green
regions are excluded by $b \to s \gamma$ and the LEP bound,
respectively.  Here we take the CP-odd Higgs mass parameter as
$m_A=1~{\rm TeV}$, $\mu=200~{\rm GeV}$ and $\tan \beta=10$. As for
gaugino masses, gluiono mass $M_3$ is taken to be $750~{\rm GeV}$ and
the GUT relation is assumed for the rest. 
The messenger scale is set to be $10^5$ GeV. The top quark mass is taken as $m_t=173.2$ GeV.
The Higgs pole mass is
calculated by using FeynHiggs~\cite{Hahn:2010te} and the $\delta
m_{H_u}^2$ is evaluated by solving renormalization group equations in
SOFTSUSY package~\cite{Allanach:2001kg}. The branching ration of $b
\to s \gamma$ is calculated by SusyBSG~\cite{Degrassi:2007kj}. In the
region where stops are light and $A_t$ is large, chargino
contributions are large and the SUSY contributions exceed the
constraint given in Eq.~(\ref{eq:bsg}). Therefore the large part of
parameter space in which the fine-tuning is relaxed, is excluded. It is
found that the maximum value of the fine-tuning parameter in the
allowed region is $\sim 5\%$.

In Fig.~\ref{fig:nega_mu}, we also show contours of $\Delta^{-1}$ with
different values of $\mu$ parameter.  In the left panel, we take
$\mu=-200$ GeV.\footnote{ Although a negative $\mu$ parameter is not
  favored in terms of the muon $g-2$, the SUSY contributions can be
  suppressed with sufficiently heavy sleptons.  } In this case, the
allowed region is slightly shifted to the direction in which $A_t$ is
small. However, the result does not change significantly and the level
of fine-tuning becomes slightly worse than the case with positive a
$\mu$.  In the right panel, we take $\mu=300$ GeV, where the maximum
value of $\Delta^{-1}$ is limited to $\sim 7\%$ (see
Eq.(\ref{eq:mu})). In this case, the allowed region turns out to be
merely wider than the case with $\mu=200$ GeV and the region with
$\Delta^{-1}\sim 7\%$ is allowed.  However, such a region is likely to be
exclude because the lightest stop mass is less than 90 GeV. Moreover the stop mass squared $m_{\tilde{Q}_3}^2$ is negative at the messenger scale; there exist a color breaking minimum, which is expected to be deeper than the electroweak symmetry breaking minimum as $V \sim m_{\rm soft}^2 M_{\rm mess}^2$. ($m_{\rm soft}$ is a typical scale of soft SUSY breaking parameters.)

\begin{figure}[t]
\begin{center}
\includegraphics[width=9cm]{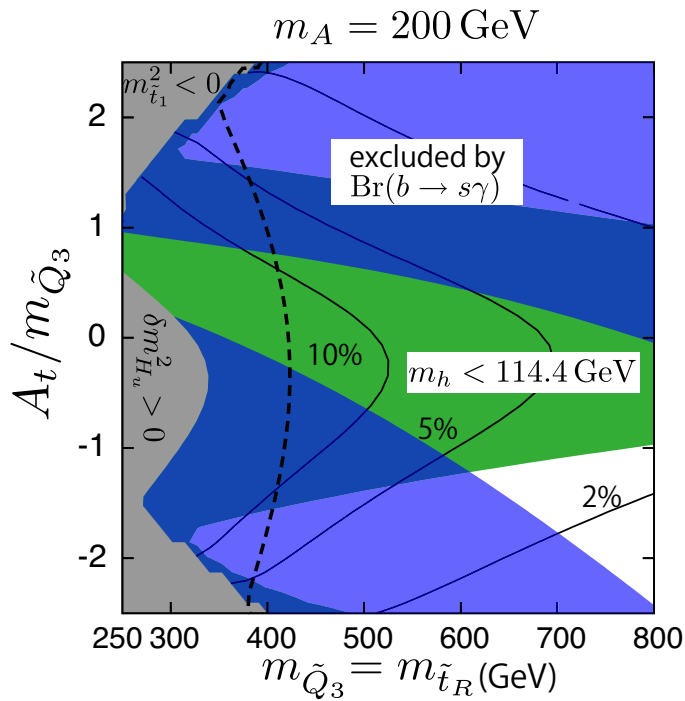}
\caption{Contours of the fine-tuning measure $\Delta^{-1}$. The CP-odd Higgs mass is taken as $m_A=200$ GeV. Other parameters are same as in   figure \ref{fig:pos_mu}.  }
\label{fig:ma200}
\end{center}
\end{figure}

We have also checked the cases where (i) $\tan\beta$ is larger
($\tan\beta=15$), (ii) $\tan\beta$ is smaller ($\tan\beta=5$) and
(iii) $m_A$ is smaller (e.g., $m_A=200$ GeV (see Fig.~\ref{fig:ma200}) and 400 GeV),
while keeping other parameters unchanged. In the case (i), it is
obvious that the chargino contributions to $b \to s \gamma$ become
larger, resulting in narrower allowed region. Therefore the level of
fine-tuning is not relaxed at all. On the other hand, in the case
(ii), although the allowed region by $b \to s \gamma$ becomes wider,
larger amount of the radiative corrections to the Higgs mass is
required to avoid the LEP constraint. This is because the tree level
Higgs mass is smaller than that with $\tan\beta=10$. In fact, the
maximum value of $\Delta^{-1}$ is smaller compared to the case with
$\tan\beta=10$. In the case (iii) the allowed region is shifted to the direction with small $A_t$ (see Fig.~\ref{fig:ma200} for example). The fine-tuning is not relaxed, even worse in this case. 


In Fig. 4 the result for larger gaugino masses is shown. The gaugino mass is taken
as $M_3=1$ TeV with satisfying the GUT relation. The fine-tuning is not ameliorated ( becomes worse ) unless the stops are tachyonic at the messenger scale.

\begin{figure}[t]
\begin{center}
\includegraphics[width=9cm]{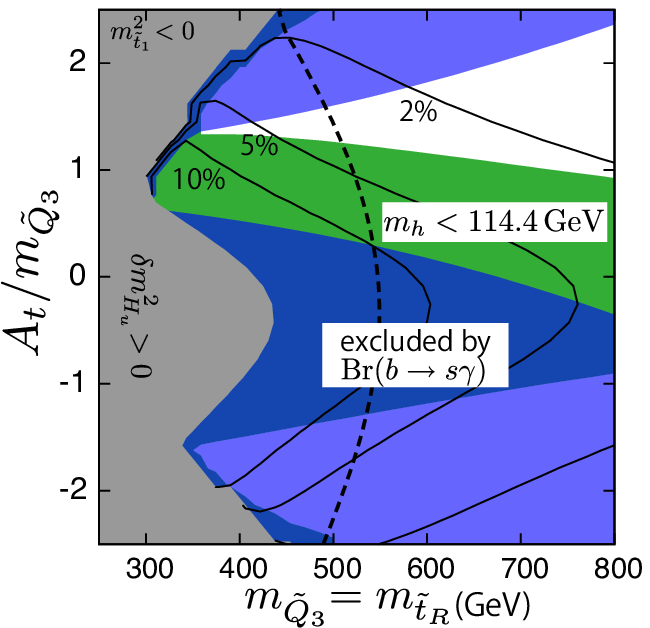}
\caption{Contours of the fine-tuning measure $\Delta^{-1}$. The parameters are same as in figure \ref{fig:pos_mu} except for gaugino masses as $M_3=1$ TeV. The Bino and Wino masses are set with the GUT relation. }
\label{fig:mgl1000}
\end{center}
\end{figure}

It is noted that we assume the GUT relation among the gaugino masses
in this work. Although we could not find a region with larger
$\Delta^{-1}$ due to $b \to s \gamma$ constraints,\footnote
{In mirage mediation~\cite{Choi:2005uz,Choi:2005hd} scenario where gaugino masses are unified at
    TeV scale, it is shown that
    $\Delta^{-1}>20\%$ is possible in the parameter region which is
    consistent with $b\rightarrow s \gamma$ observation~\cite{Kitano:2005ew}.}
 it is possible to find a ``less'' fine-tuned region with an elaborate choice of $m_A$
and $M_2$, once the GUT relation is relaxed;
 the different
contributions to $b \to s \gamma$ can cancel each other, which is yet
another tuning.
  
Finally we comment on Higgs phenomenology at the LHC. In the allowed
region, we also calculated Higgs production rate at the LHC and its
branching ratio of each decay mode.  At the LHC, gluon fusion is the
main production process. It turns out that the Higgs production rate
is almost unchanged compared to the SM value.  The Higgs decay
property is also similar to that in the SM.  The decay mode
$h\rightarrow \gamma \gamma$ is especially important for the discovery
of Higgs boson, as well as the determination of its mass, in the light
Higgs scenario.  We found a few \% difference in the decay rate.  The
channels in which Higgs decaying to $WW$ and $ZZ$, on the other hand,
are reduced by up to about $10\%$.

\section{Conclusion}
In this letter we have studied the level of fine-tuning in the Higgs
sector, considering the constraints from the observation of the
branching ratio for $b \to s \gamma$ in the MSSM. While light stops
are favored by the relaxation of the fine-tuning, it would predict the
large branching ratio which significantly deviates from the
experimental value. It is found that the parameter region where the
fine-tuning measure is larger than 5\% (10\%) is excluded by $b \to s
\gamma$ constraints even for low messenger scale as $10^5$ GeV ($10^4$
GeV), {assuming the GUT relation among gaugino masses.}  Therefore,
being consistent with the present experiments, realization of the
natural supersymmetry is difficult. Note that although the light stops
may avoid the constraints from SUSY searches in the cases that
$R$-parity is violated, our result can be applied even for such a case.

\section*{Acknowledgment}
We thank Kazuhiro Tobe for useful discussion.  N.Y. would like to
thank Caltech Particle Theory group where this work has been carried
out.  This work is supported by Grant-in-Aid for Scientific research
from the Ministry of Education, Science, Sports, and Culture (MEXT),
Japan, No. 22-7585(N.Y.).  This work was also supported in part by the
U.S. Department of Energy under contract No. DE-FG02-92ER40701, and by
the Gordon and Betty Moore Foundation (K.I.).

\providecommand{\href}[2]{#2}\begingroup\raggedright\endgroup

\end{document}